\documentclass[journal=jacsat,manuscript=article]{achemso}

\usepackage{chemformula} 
\usepackage[T1]{fontenc} 
\RequirePackage{graphicx,amsmath,amssymb,bm}
\RequirePackage{hyperref}

\newcommand{\Fref}[1]{Fig.~\ref{#1}}

\renewcommand{\eqref}[1]{eq.~(\ref{#1})}

\author{Dario A. Bahamon}
\affiliation{School of Engineering, Mackenzie Presbyterian University, S\~ao Paulo - 01302-907, Brazil}
\alsoaffiliation{MackGraphe Graphene and Nanomaterials Research Institute, Mackenzie Presbyterian University, S\~ao
Paulo -01302-907, Brazil}
\alsoaffiliation{Departamento de  Teor\'ia y Simulaci\'on de Materiales, Instituto  de Ciencias de Materiales de Madrid, CSIC, E-28049, Madrid, Spain}
\email{dario.bahamon@mackenzie.br}
\author{Guillermo G\'omez-Santos}
\affiliation{Departamento de F\'{\i}sica  de la Materia Condensada,
Instituto Nicolás Cabrera and Condensed Matter Physics Center (IFIMAC),
Universidad Autónoma de Madrid, E-28049 Madrid, Spain}
\email{guillermo.gomez@uam.es}
\author{Dmitri K. Efetov}
\affiliation{Fakult\"at f\"ur Physik, Ludwig-Maximilians-Universit\"at, Schellingstrasse 4, M\"unchen 80799, Germany}
\alsoaffiliation{Munich Center for Quantum Science and Technology (MCQST), M\"unchen, Germany}
\email{dmitri.efetov@lmu.de}
\author{Tobias Stauber}
\affiliation{Departamento de  Teor\'ia y Simulaci\'on de Materiales, Instituto  de Ciencias de Materiales de Madrid, CSIC, E-28049, Madrid, Spain}
\email{tobias.stauber@csic.es}

\title{Chirality probe of twisted bilayer graphene in the linear transport regime}


\makeatletter
\newcommand*{\forcekeywords}{
  \acs@keywords@print
  \let\acs@keywords@print\relax
}
\makeatother

\begin{document}



\begin{abstract}
We propose minimal transport experiments in the coherent regime that can probe the chirality of twisted moir\'e structures. We show that only with a third contact and in the presence of an in-plane magnetic field (or other time-reversal symmetry breaking effect), a chiral system may display non-reciprocal transport in the linear regime. We then propose to use the third lead as a voltage probe and show that opposite enantiomers give rise to different voltage drops on the third lead. Additionally, in the scenario of layer-discriminating contacts, the third lead can serve as a current probe, capable of detecting different handedness even in the absence of a magnetic field. In a complementary configuration, applying opposite voltages on the two layers of the third leads gives rise to a chiral (super)current in the absence of a source-drain voltage whose direction is determined by its chirality. 
\end{abstract}

\forcekeywords 

\section{Introduction}

\noindent The idea of chirality permeates many  branches of science, \cite{wagniere2007chirality,Salam:1991aa,Wilczek} and also in condensed matter physics the concept is used to describe electronic properties in reciprocal and real space. In reciprocal space, chirality defines the handedness of the spin of the electron to its momentum,\cite{Wang:2017aa,RevModPhys.88.035005,felser2022topology} in real space, chirality depicts the handedness of molecules and solids that cannot be superimposed onto their mirror images.\cite{wagniere2007chirality,Chiralspintronics} Most generally, chirality always emerges when discrete symmetries such as refection, time-reversal or particle-hole symmetry are broken. 

Independent of the setting, chiral systems offer opportunities to observe new phenomena as well as challenges regarding their detection. \cite{felser2022topology,Hasan:2021aa,Chiralspintronics} The signature of topological insulators, e.g., is the existence of chiral dissipationless states at the boundaries of the sample,\cite{Wang:2017aa,felser2022topology,Hasan:2021aa} and the electrical detection of these states requires non-local transport measurements. \cite{RES_VALENZUELA} On the real space side, chiral molecules spin polarize the electric current passing through them,\cite{10.1021/acsnano.1c01347}  and the chiral-induced spin selectivity (CISS) effect is detected in a two terminal configuration as a non-linear I-V characteristic.\cite{Chiralspintronics} Also, electrical magneto-chiral anisotropy in a classical four 
terminal configuration has been observed, where the longitudinal enantio-selective 
magneto-resistance is non-linear in the current.\cite{MChPop,PhysRevLett.122.057206,Yokouchi:2017aa} Finally, in non-centrosymmetric materials, chirality manifests itself in various non-reciprocal response phenomena \cite{NRTokura,NRCheong,Ando20,RevMChA,MChDie} where the resistance depends linearly on the current causing a non-linear voltage drop in a two terminal setup. \cite{rikken2001electrical,chiralnt,NRvdw}

Van der Waals moir\'e materials offer new opportunities to engineer certain geometric structures that can lead to novel properties.\cite{NRvdw,PhysRevB.104.075407,Scammell_2022} 
For example, twisted bilayer graphene\cite{Lopes07,Suarez10,Bistritzer11,Moon12,PhysRevLett.108.216802,PhysRevB.93.035452} (TBG), which has recently attracted great attention due to discovery of novel electronic phases,\cite{Cao18b,Yankowitz19,Codecido19,Shen19,Lu19} can be rotated clock- or anti-clockwise. Due to the finite interlayer separation, TBG heterostructures with opposite twist angles can only be superimposed onto each other after performing a mirror reflection with respect to the $xy$-plane. Thus, TBG is intrinsically chiral which has been experimentally demonstrated by observing optical dichroism without breaking time-reversal symmetry.\cite{Kim16} The effect becomes largest for frequencies which induce transitions between states that are maximally delocalized between the two layers such that the misalignment between layers is most effective.\cite{Suarez_Morell_2017,Stauber18} But also in the dc-limit, the chirality of TBG is manifested by intrinsic magnetic-electric coupling.\cite{Stauber18b,PhysRevB.100.125418,Stauber20C,Stauber20,Lin20,HeWenYo20,Zuber21,Margetis21,Timmel21,Huang22,Antebi22,Dutta23,Stauber23,Zhai23} In the  dc-limit with broken time reversal symmetry, an electrical magnetochiral anisotropy is anticipated to emerge. \cite{liu2021chirality} Nevertheless, in conventional transport experiments where the net current in voltage leads is maintained at zero,\cite{SECButtiker,Datta} the influence of chirality has not been a significant factor thus far.
 
In this work, we precisely fill in this gap by investigating electronic transport through TBG in the linear regime within the Landauer-B\"uttiker formalism. Using general symmetry arguments, we point out that in order to distinguish different enantiomers, i.e., samples with twist angles $\theta$ and $-\theta$, respectively, it is crucial to have three leads. This is contrary to the zero-field superconducting diode effect where two leads are sufficient.\cite{Ando20,Baumgartner22,Bauriedl22,Hou22,Yuan22,Lin22,DiezMerida23,Picoli23} The third lead can now either be used as a voltage probe that detects the chirality if a magnetic field in applied parallel to the layer. Or, in the case of layer-discriminating contacts, it can be used as current probe even in the absence of a magnetic field. We thus show that it is not necessary to break time-reversal symmetry with a magnetic field in order to observe the layer-discriminating transverse currents  effect in a typical transport experiment.

\section{Landauer formalism and general symmetries}

\begin{center}
\begin{figure}[!h]
\scalebox{0.5}{\includegraphics{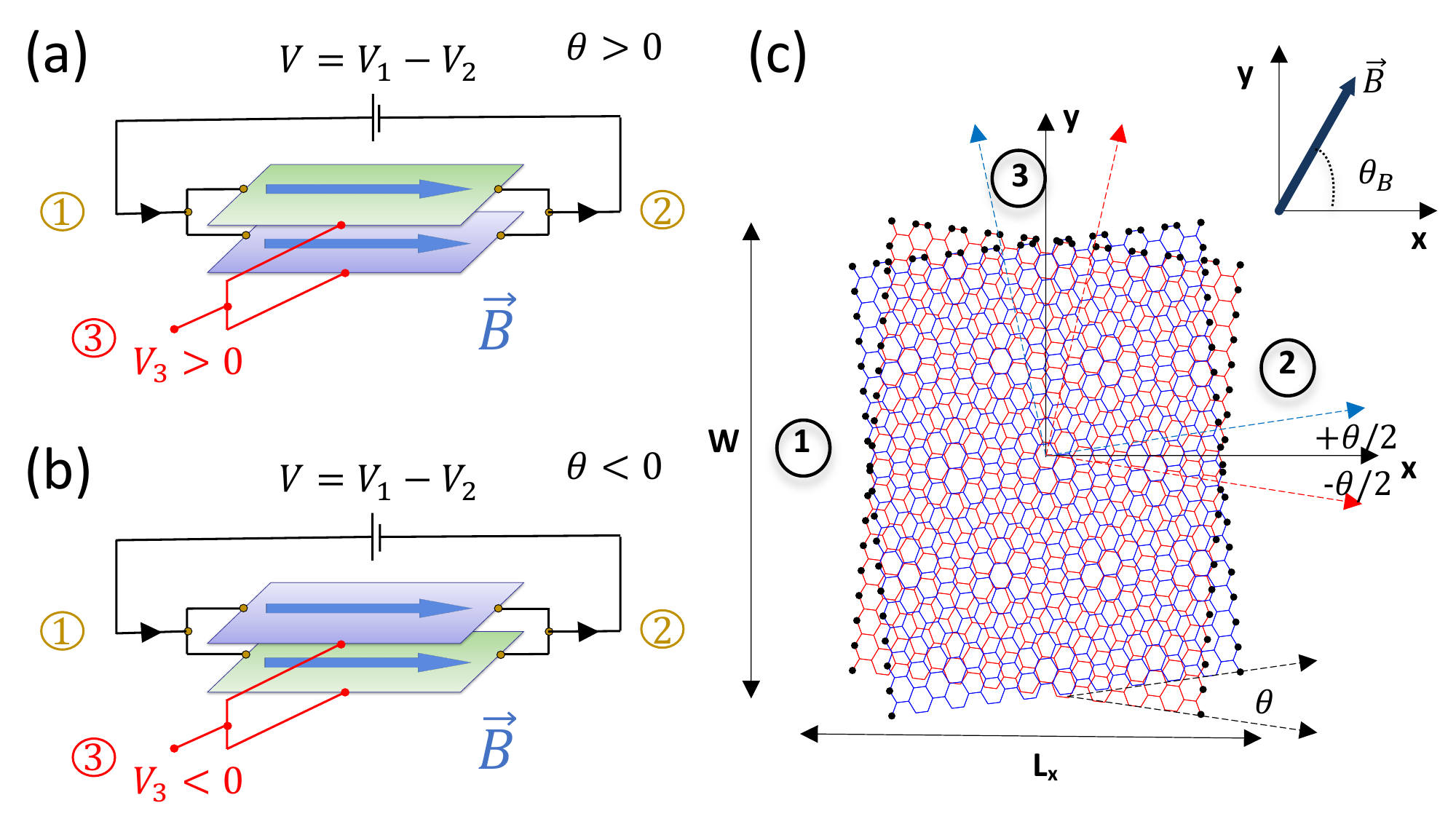}}
\caption{ (Color online) {(a) and (b): Proposed setup for voltage probe detection of  chirality  in the presence of a magnetic field (${\bm B} = B\,{\bm e}_x$). The reading of the voltage $V_3$ is opposite for the opposite chiralities depicted}. The Green color stands for a layer twisted by a positive angle, while the violet  means a negative twisted  angle  layer.  (c) Schematic representation of the twisted bilayer graphene (TBG)  three-terminal device, the  yellow and red dots represent the atomic sites in contact with leads 1, 2 and 3 respectively.}
\label{fig:device} 
\end{figure}
\end{center}

We consider two systems with opposite twist angles $\pm\theta$ or chiralities, related to each other by a mirror symmetry with respect to a horizontal plane midway between the layers. Notice that in this symmetry operation we include the leads, which we consider, for the moment, equally coupled to both layers.  The reason for this  layer-symmetric attachment of leads is twofold:  first, hopefully easier   experimental realization, and second,   simplified chiral analysis for, otherwise,  a right lead attached to the bottom layer in the $\theta$ flake would become a right lead, but in the top layer for the mirror flake with opposite chirality $-\theta$. Let us also assume that there are $N$ leads attached to the system, and that the current $I_p$ in lead $p$ is \cite{Datta}

\begin{align}
\label{Ohm}
I_p = \sum_qG_{pq}[V_p -V_q],
\end{align}

\noindent where $V_q$ is the potential of lead $q$ over ground level and the conductance $G_{pq} = \frac{2e^2}{h} \overline{T}_{pq}$ is proportional to the total transmission $ \overline{T}_{pq}$ between lead $q$  lead $p$. Because of the inherent symmetries of the transmission matrices, the following reciprocity relation holds\cite{Datta} 

\begin{align}
\label{eq:recip}
G_{pq}(\theta,{\bm B}) = G_{qp}(\theta, -{\bm B})\;,
\end{align}

\noindent where  $\bm B$ is the magnetic field, see the Supplementary Information (SI). In our case, it will denote an in-plane magnetic field.

We now perform a z-reflection by mapping $z$ to $-z$ for each piece of matter, including sources if external fields are present. This is the composition of space-inversion, $(x,y,z) \to (-x,-y,-z)$, followed by a $\pi$-rotation around the z-axis, see Fig. \ref{fig:device}-(c), where the origin is placed in the center and midway between the planes. 
By this, one maps $\theta\to-\theta$, $p\to p$, $q\to q$, and, crucially important here, ${\bm B} \to -{\bm B}$, i.e., ${\bm B}$ is an axial vector, first unchanged by space-inversion but later changing sign upon the $\pi$-rotation around the z-axis. In twisted systems, we thus also have the {\it chiral} reciprocity relation
\begin{align}
\label{Treflection}
G_{pq}(\theta, {\bm B}) = G_{pq}(-\theta, -{\bm B})\;,
\end{align}
valid for in-plane magnetic fields and layer-symmetric attachments. This correlation shows that there is no way to detect the sign of the chirality without magnetic fields in twisted arrangements with layer-symmetric leads, because $G_{pq}(\theta, {\bm B} = {\bm 0}) = G_{pq} (-\theta, {\bm B} = {\bm 0})$. On the other hand, the correspondence expressed in Eq. (\ref{Treflection}) implies a locking of chirality and field. 

To exploit this relation, let us now fix the direction, ${\bm e}_{\bm n},$ of the  in-plane magnetic field, ${\bm B}=B{\bm e}_{\bm n}$ and expand the conductance to linear order in $B$,

\begin{align}
\label{expansionB}
G_{pq} (\theta, B) = G_{pq} (\theta, 0) + \Delta_{pq} (\theta)B\;,\\
G_{qp}(\theta, B) = G_{qp}(\theta, 0) + \Delta_{qp}(\theta)B\;.
\end{align}

\noindent In conjunction with Eqs. (\ref{eq:recip}) and (\ref{Treflection}) and knowing, as stated earlier,  that $G_{pq} (\theta, 0) = G_{qp}(\theta, 0) = G_{ qp}(-\theta, 0)$ we arrive at 

\begin{align}
\Delta_{pq} (\theta) = -\Delta_{qp}(\theta) = -\Delta_{pq} (-\theta)\;.
\end{align}

\noindent The above equation indicates an electrical mechanism to detect chirality of twisted devices through conductance measurements  even with layer-symmetric leads. 

\section{ Including the third lead as voltage probe for chirality}

Let us now focus on the minimal system that can detect chiral properties in the linear regime. The device with three leads is depicted in Fig. \ref{fig:device}(a)-(b). All leads couple equally to both layers, as previously assumed and lead 1 and 2 have equal number of channels. In the absence of a magnetic field, the currents can be obtained from Eq. \ref{Ohm}

\begin{equation}\label{MatI}
\left(\begin{matrix} I_1\\I_2\\I_3\end{matrix}\right)= \left(\begin{matrix} G_{12}^0+G_{13}^0 &-G_{12}^0&-G_{13}^0\\-G_{12}^0&G_{12}^0+G_{13}^0&-G_{13}^0\\-G_{13}^0&-G_{13}^0& 2G_{13}^0\end{matrix}\right)\left(\begin{matrix} V_1\\V_2\\V_3\end{matrix}\right)\;,
\end{equation}

\noindent where it is defined that $G_{pq}(\pm \theta,{\bm B} = {\bm 0}) = G_{pq}^0 = G_{qp}^0$ as well as assumed, based on the symmetry of the problem, that $G_{13}^0 = G_{23}^0$.  When ${\bm B} \neq {\bm 0}$, the additional  contribution to the conductance matrix  is:

\begin{equation}\label{deltaI}
\left(\begin{matrix} \delta I_1\\\delta I_2\\\delta I_3\end{matrix}\right)= \left(\begin{matrix} 0&-\delta G&+\delta G\\+\delta G&0&-\delta G\\-\delta G&+\delta G&0\end{matrix}\right)\left(\begin{matrix} V_1 \\ V_2 \\ V_3 \end{matrix}\right)\;,
\end{equation}

\noindent where the corrections are characterized by a single parameter to linear order in $|{\bm B}|$, $\delta G = \Delta(\theta)B$. This is a consequence of the conductance sum rule $\sum_{q\neq p}G_{pq}(+{\bm B}) =  \sum_{q\neq p}G_{pq}(-{\bm B})$ which implies that  $\sum_{q\neq p}\Delta_{pq} = 0$. Now, we can  use the third lead as a voltage probe, imposing $I_3=0$ while applying a source-drain voltage drop,  $V_1=V/2,~V_2=-V/2$. Then to linear order in the $B$-field, the voltage probe, $V_3$, yields 

\begin{align}
\frac{V_3}{V}=\frac{\delta G}{2G_{13}^0} = \frac{\Delta(\theta)}{2G_{13}^0}B .
\label{eq:V3}
\end{align}

\noindent Previously, it was demonstrated that $\Delta(\theta) = -\Delta(-\theta)$, this ensures that the voltage probe becomes also a probe of the chirality sign ($V_3(\theta) = -V_3(-\theta)$). 

\subsection{Numerical implementation}  Having established the basic relations that allow the  detection of handedness  through electric measurements, our attention turns to their numerical calculation  in a  TBG region of dimensions $W \times L_x$, where both $W$ and $L_x$ are set to 50 nm (the system is sketched in \Fref{fig:device}(e)). The computation is performed employing the tight-binding model and Green's functions. It's crucial to note that we opt for multiple neighbors for each site to replicate the characteristics of TBG. As a consequence, the charge neutrality point (CNP) emerges at approximately $E \approx 0.2961t_0$, where $t_0$ represents the nearest neighbor hopping. Consequently, all numerical outcomes presented in this study are offset by this value and $E_F = E - 0.2961t_0$. For detail, we refer to the supplemental information. There, we also show that the reciprocity relations as delineated in Eqs. (\ref{eq:recip}) and (\ref{Treflection}) are correctly implemented in our numerical calculations. 
\begin{center}
\begin{figure}[h!]
\scalebox{0.98}{\includegraphics{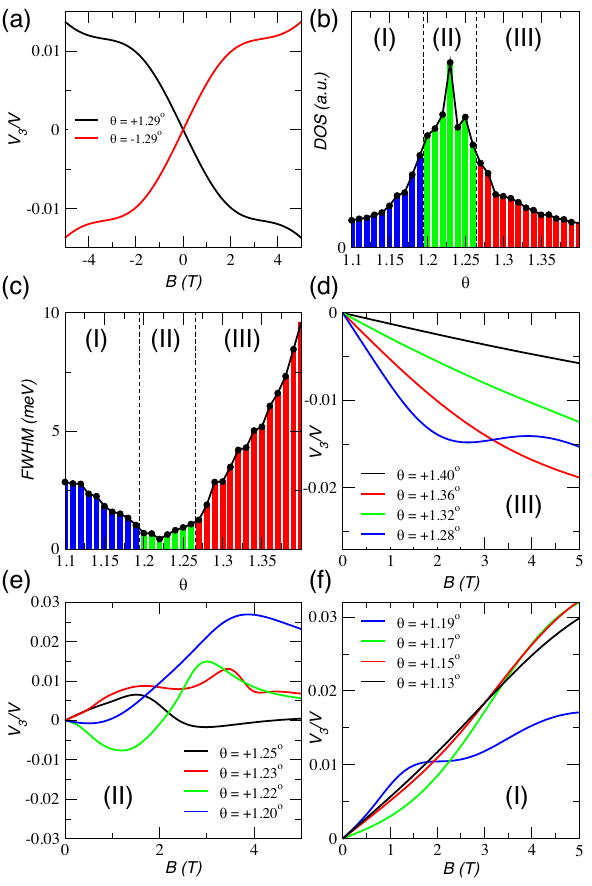}}
\caption{ (Color online) (a), (d)-(f) $V_3$ for the different twist angles as function of ${\bm B}$ parallel to the $x$-axis.  (b) Value and (c) full width at half maximum (FWHM) of the DOS peak without magnetic field. The Roman numerals I, II, and III, along with the shaded regions, serve as visual guides to differentiate various chirality behaviors. In the panels where $V_3/V$ is presented, the Fermi energy is set to zero, $E_F = 0$. 
}
\label{fig:V4DOS} 
\end{figure}
\end{center}

\section{Detecting chirality by voltage probe with a magnetic field}
We now consider the case where $V_1=V/2,~V_2=-V/2~\text{and}~I_3=0$. Using this condition, $V_3/V = (G_{31} - G_{32})/(2~G_{31} + 2~G_{32})$ can be calculated as function of the magnetic field for different twist angles and energies.  For $\theta = \pm 1.29^\circ$ and Fermi energy $E_F = 0$ meV,  Fig. \ref{fig:V4DOS}(a) shows that $ V_3/V $ is linear in B with opposite slope for opposite twist angles, confirming the prediction of  Eq. (\ref{eq:V3}) that lead 3 becomes also a probe for chirality and that quantum transport is sensitive to chirality. However,  it is not possible to relate in advance the sign nor the strength of the linear field-dependence to a certain twist angle.  This is similar to  previous observations for the infinite system,\cite{Stauber18,Stauber18b,Stauber20C}  where the chiral component of the conductivity was shown to exhibit  highly non-monotonous behavior with filling factor and twist angle.

For large angles, the effect of the handedness of the junction on transport is negligible and $V_3/V_1 \approx 0$. This already indicates that the effect is mainly electronically and not configurationally driven, for both layers are highly decoupled for large  angles. For small angles, the density of states (DOS) is presented in Fig. \ref{fig:V4DOS}(b)-(c) in terms of the the largest value as well as the full width at half maximum (FWHM) of the main peak. From there, we infer that the magic angle is located around $\sim1.23^{\circ}$ where the highest and narrowest DOS peak appears and serves to diagnose the magic-angle regime. 

Additionally, three regions are shaded and identified by the Roman numbers I, II and III conforming  to the behavior of $V_3$. For $\theta > 1.26^{\circ}$ (region III in red),  a linear-in-$B$ regime is perfectly defined with negative slopes for positive twist angles as shown in Fig. \ref{fig:V4DOS}(d). In the blue shaded region I ($\theta < 1.20^{\circ}$), the signal of the linear response is inverted and positive slopes appear for positive twist angles; characteristic lineshapes of $V_3$ for angles in these region are presented in Fig. \ref{fig:V4DOS}(f). 

Around the magic angle, the linear relation on the magnetic field becomes weaker and the non-linear $B$-dependence dominate the response as shown in Fig. \ref{fig:V4DOS}(e). However, even for twist angles in this region, the chirality obeys the symmetry relation $V_3(+\theta) = -V_3(-\theta)$) as discussed in the SI, where also the effect of different coupling strengths to lead 3 are presented. 

Let us finally emphasize that our approach utilizes the Landauer-B\"uttiker formalism within the linear regime. In this regime, the conductance $G$ is computed in equilibrium, thus avoiding any dependence on current density. This distinction ensures that in our setup, the relationship between the voltage probe reading, $V_3$, and the  source-drain voltage (and  current)  remains linear.

\section{Detecting chirality by current probe without a magnetic field} 
\begin{center}
\begin{figure}[!h]
\scalebox{0.5}{\includegraphics{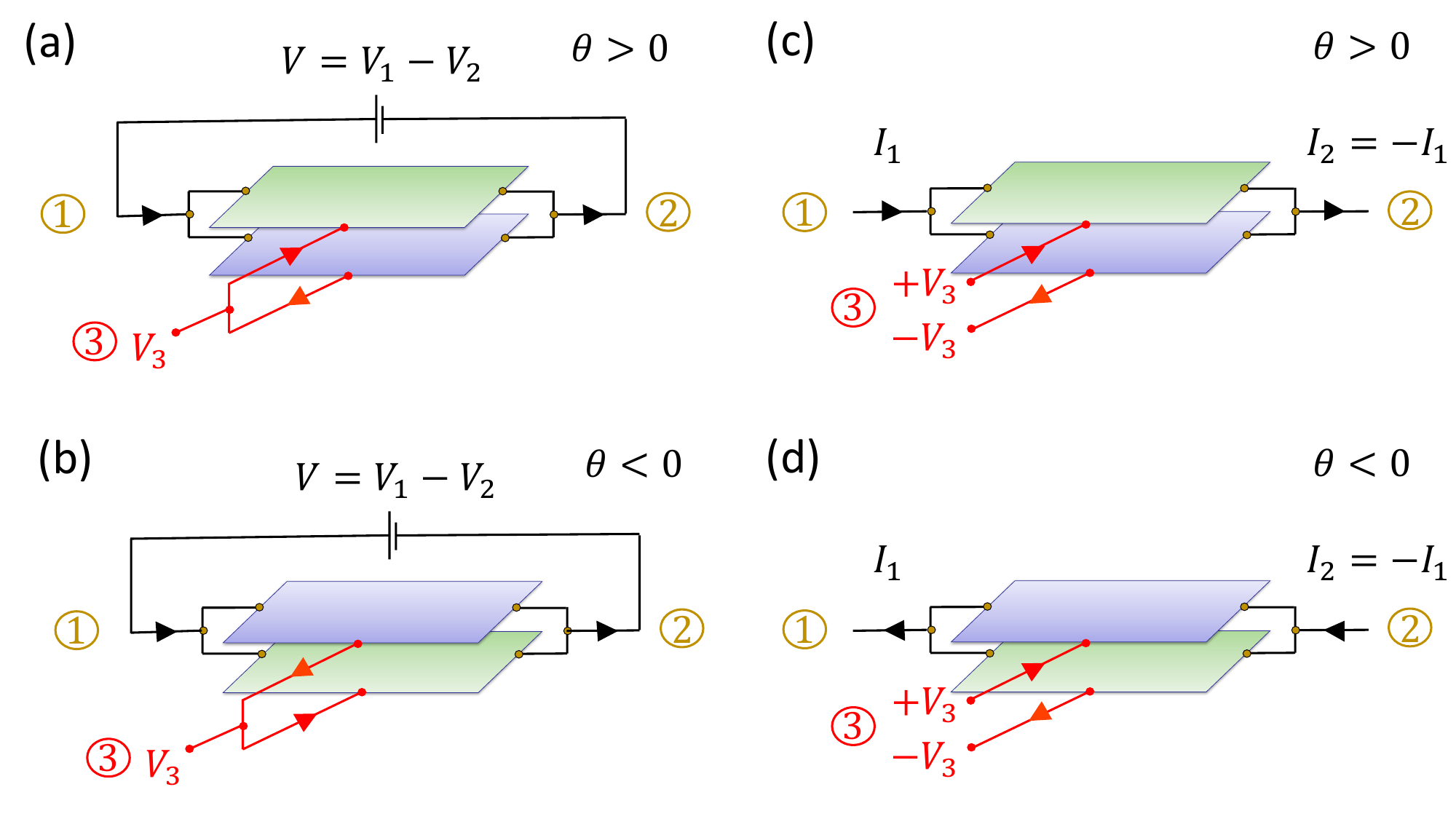}}
\caption{ (Color online) (a) and (b): Proposed setup for chirality detection by current probe  without magnetic field. Top and bottom transverse currents reverse direction for the opposite chiralities depicted. The Green color stands for a layer twisted by a positive angle, while the violet  means a negative twisted  angle  layer. The detection of chirality by current probe requires now four leads (1, 2 3t and 3b), because the layer index top and bottom of lead 3 is now discriminated. (c) and (d): Alternative set-up showing the same electro-magnetic coupling without magnetic field. Now, there is no source-drain voltage, i.e., the potential $V_1$ and $V_2$ are equal, but a potential difference between the top and the bottom layer induces a current whose direction depends on the chirality.}
\label{fig:device2} 
\end{figure}
\end{center}
Let us now discuss how to detect the chirality of a system without a magnetic field. In the infinite TBG, the transverse conductivity is equal in magnitude, but of opposite direction with respect to the two layers due to basic symmetry constrains. \cite{Stauber18}  
\subsection{Inducing a transverse current}
In the setup proposed in Figs \ref{fig:device2}(a)-(b), we  expect to see layer-discriminating transverse currents without the presence of a magnetic field as they are allowed by  the same symmetry principles. This implies that a source-drain voltage $V = V_1 - V_2$ between lead 1 and lead 2 will be accompanied by transverse currents in both layers, flowing in opposite directions. As $V_{3t} = V_{3b}$, one can interpret this as a vertical "supercurrent" flowing from one layer to the other. In an infinite system, this effect implies that a net current in the $x$ direction is accompanied by layer-opposite currents in the $y$ direction, which can be thought of as an in-plane magnetic moment along the net flow, a hallmark of chirality\cite{Stauber18,Stauber18b,bahamon2020emergent,Furukawa:2017aa,PhysRevLett.124.166602}.

\begin{center}
\begin{figure}[h!]
\scalebox{0.98}{\includegraphics{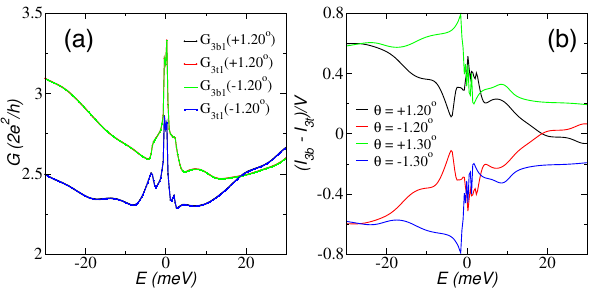}}
\caption{ (Color online) (a) Conductance (in units of $2e^2/h$) from lead 1 to lead 3 bottom layer ($G_{3b\,1}$) and from lead 1 to lead 3 top layer ($G_{3t\,1}$) for $\theta = \pm 1.20^{\circ}$. (b) $G_{3t\,1}-G_{3b\,1} = (I_{3b} - I_{3t})/V $ (in units of $2e^2/h$) for $\theta = \pm 1.20^{\circ},\pm 1.30^{\circ}$. From the formalism sign conventions, the finite value of the plotted magnitude implies currents of equal magnitude, $|I_{3b}| = |I_{3t}|$, running opposite in opposite leads, and reversing direction upon chirality reversal, as depicted in Fig. \ref{fig:device2}(a)-(b).}
\label{fig:Icapa} 
\end{figure}
\end{center}

We need first to generalize Eq. (\ref{Treflection}) by including the layer index via $q\to q\nu$ with $\nu=t,b$. Performing the reflection and assuming the same coupling between the two layers as schematically shown in Figs. \ref{fig:device2}(a)-(b), we then have

\begin{align}
\label{chiralityWithoutB}
G_{q\nu,p\mu}(\theta)=G_{q{\bar \nu},p{\bar{\mu}}}(-\theta)\;,  
\end{align}

\noindent where $\bar \nu, \bar \mu$ denote the opposite layer of $\nu,\mu$. We can now extend  Eq. (\ref{MatI}) to effectively four leads: leads one and two remain layer-symmetric as before, while the original third lead splits into top ($3t$) and bottom ($3b$), see  \Fref{fig:device}(c)-(d) and SI for reference.  With the above in mind, in Fig. \ref{fig:Icapa}(a) we show the conductance from lead 1 to  lead 3 in the top layer, $G_{3t 1}$, and the conductance from lead 1 to the lead 3 in the bottom layer, $G_{3b 1}$. Since the contacts are symmetric, there  would be no difference in the two conductances  if the system were achiral  without a magnetic field. However, a clear difference is seen giving rise to a chiral current probe in the absence of a magnetic field. Note that reversing the angle maps the conductance $G_{3b 1}(\theta)$ to $G_{3t 1}(-\theta)$ and vice versa, as dictated by the general symmetry relations of Eq. (\ref{chiralityWithoutB}).  Notice that there is no need for dealing separately with lead 2 for, in our geometry, $ G_{3t\, 2} = G_{3b \,1}$ and $ G_{3t \,1} = G_{3b\, 2} $.

Now, we can consider the arrangement schematized in Figs. \ref{fig:device2}(a)-(b), $V_1=V/2$, $V_2=-V/2$, and  $V_{3b}=V_{3t}=0$, in which no {\em net} current flows between system and the two $3t$ and $3b$ reservoirs: $I_{3b}+I_{3t} = 0$.  Yet, a finite current from the upper lead to the lower lead can then be deduced from the finite value of $\delta G(\theta) = (I_{3b}-I_{3t})/(V_1 - V_2) = G_{3t\, 1}(\theta) - G_{3b \,1}(\theta)$. Remember that, from the sign conventions of the formalism, this implies  currents of equal magnitude, $|I_{3b}| = |I_{3t}|$, running in opposite directions in top and bottom leads, despite the terminals $3t$ and $3b$ having equal potentials, as illustrated in Figs. \ref{fig:device}(c)-(d). Reversing the chirality reverses the layer currents, $I_{3b(t)}(\theta) = -I_{3b(t)}(-\theta)$, therefore the experimental detection of these currents becomes a  probe  for chirality without magnetic field ($\delta G(\theta) = -\delta G(-\theta)$) as shown in Fig. \ref{fig:Icapa}(b). 

The detection of this vertical "supercurrent" can be difficult with real, lossy leads. Alternatively, one could use the top and bottom third leads as independent voltage probes,
whose different readings would then reveal the chirality, as explained in the SI.

\subsection{Inducing a longitudinal current}
A complementary configuration based on the same symmetry principles is presented in Figs \ref{fig:device2}(c)-(d), where the same phenomenon also gives rise to a "supercurrent" from reservoirs 1 to 2  without any voltage drop between them,  $V_1 = V_2=0$, when now the transverse currents are driven by a corresponding voltage drop, $V_{3t} = -V_{3b} =V$. 

A finite current of magnitude $|I_1|=I_2|$ from the lead 1 to lead 2 (or viceversa), can then be deduced from the finite value of $\delta G(\theta) = (I_{1}-I_{2})/(V_{3t} - V_{3b}) = G_{3t\, 1}(\theta) - G_{3b \,1}(\theta)$. Note that $\delta G(\theta)$ is given by the same expression as in the previous subsection and which is shown in Fig. \ref{fig:Icapa}(b). The electro-magnetic coupling is thus the same in both cases as already discussed in Ref. \cite{Stauber18}. Further discussion can be found in the SI, where it is also shown how to,
instead, use terminals 1 and 2 as voltage probes to bypass the difficulties of lossy leads in the "supercurrent" detection. 

\section{Summary and conclusions.} In summary, we have proposed two linear transport experiments that can detect the intrinsic handedness of chiral systems. First, exact symmetry arguments were provided to show that a voltage probe can become a chirality probe in the presence of a in-plane magnetic field for external leads that couple symmetrically to both layers. Secondly, we also demonstrated that a current probe  discriminating between the two layers can become a chirality probe even in the absence of a magnetic field. Different enantiomers can thus be distinguished by minimal transport experiments that will hopefully shed more light on this intriguing symmetry, also present in various organic molecules. 

Using the Landauer-B\"uttiker formalism, the voltage reading of the third lead and the layer-discriminating, opposite currents, between system and  third reservoir(s) were explicitly calculated for a finite sample of twisted bilayer graphene (TBG), confirming our predictions.  
In both situations, when approaching the magic angle, there is change of chirality measured through electronic means (voltage and current probes) that does not correspond to an inversion of the actual twist angle between the layers.\cite{Stauber20C} 
In the case of layer-discriminating leads, we showed that, in the presence of chirality,  a net current flow between source and drain reservoirs is accompanied by transverse currents, opposite in each layer. This provides a realization in a Landauer-B\"uttiker scenario, more accessible experimentally, of  predictions previously made\cite{Stauber18,Stauber18b,Stauber20C}  on the basis of linear response for infinite systems, that in-plane magnetic moments should accompany net current flows in chiral bilayers.

Our one-particle formalism does not allow for symmetry broken ground-states. However, for a time-reversal symmetry (TRS) broken ground-state $|GS\rangle$ we expect an alternative {\it chiral} reciprocity relation even in the {\it absence} of a magnetic field, involving only the ground-state $G_{pq}(\theta, |GS\rangle) = G_{pq}(-\theta, \mathcal{T}|GS\rangle)$ where $\mathcal{T}$ denotes a the anti-unitary time-reversal operator. This relation allows for the detection of a TRS broken ground-state by measuring a non-zero voltage at the third lead at ${\bf B}=0$.

Our discussion should also be relevant for other chiral systems, e.g., for those displaying the planar Hall effect.\cite{Burkov17,Nandy17} Then, the continuous variable $\theta$ denoting the twist angle is simply replaced by a discrete variable $\chi=\pm1$ denoting the different enantiomers. Even the emergence of a magic angle seems to be more general since it can be observed in Weyl semimetals, too.\cite{Pixley18} Let us finally note that the third lead can also be realized by a local probe, especially in the layer-discriminating case. Real-space mapping of the handedness and thus small angle-deviations around the magic angle should be detectable.

\begin{acknowledgement}

This work has been supported by project nos. PGC2018-096955-B-C42,
PID2020-113164GB-i00, and CEX2018-000805-M financed by MCIN/ AEI/
10.13039/501100011033 as well as by the CSIC Research Platform on Quantum Technologies PTI-001. DAB acknowledges support from the Brazilian Nanocarbon Institute of Science and Technology (INCT/Nanocarbon), CAPES-PRINT (grants nos. 88887.310281/2018-00 and 88887.899997$/$2023-00),  CNPq (309835/2021-6) and Mackpesquisa. DAB is also grateful to the hospitality of the ICMM where this work was completed.

\end{acknowledgement}

\newpage
{\huge\bf Supplementary Information}

\section{Green´s functions and tight binding method }

Employing   the Green's functions we have  $G_{pq}=\frac{2e^2}{h}\text{Tr}[\Gamma_p \mathcal{G} \Gamma_q \mathcal{G}^{\dagger}]$, where $\mathcal{G} = [E-H_{TBG}-\Sigma_1-\Sigma_2 - \Sigma_3]^{-1}$ is the Green's function of the central region, $\Sigma_{1(2)(3)}$ are the self-energies of the leads  and $\Gamma_{1(2)(3)} = i[\Sigma_{1(2)(3)} - \Sigma_{1(2)(3)}^\dagger]$ are the  couplings of the central region to the leads. Without loss of generality we can assume a wide band model for the leads, that is a constant density of states (DOS) around the Fermi energy, which is typical in metallic contacts. In that case, the self-energy  term can be written as $\Sigma_{1(2)(3)} = -i \pi \rho|t|^2$,\cite{Datta} were $\rho$ is de DOS of the contact and $t$ is the hopping parameter between the  leads and the central twisted region. To guarantee a large number of injected modes  we set $\rho_{L(R)} = (\pi t)^{-1}$ were $t$ is the nearest neighbor hopping of graphene. \cite{bahamon2020emergent} 

The Hamiltonian of the central region ($H_{TBG}$) is   described by a tight-binding model where the hopping amplitudes  between sites $i$ and $j$, $t_{ij}(d_{ij}) =  V_{pp\sigma}(d_{ij})\cos^2(\phi) + V_{pp\pi}(d_{ij})\sin^2(\phi)$, where the bond length $d_{ij} = |{\bm d}_{ij}| = |{\bm R}_j - {\bm R}_i|$ and $\phi$ denotes the angle formed by ${\bm d}_{ij}$ and the \textit{z}-axis. The value of the inter-atomic matrix elements is a function of the bond length:\cite{Brihuega12,Moon12} $V_{pp\sigma} = V_{pp\sigma}^0 e^{-\frac{d_{ij}-d_0}{\delta}},~V_{pp\pi} = V_{pp\pi}^0 e^{-\frac{d_{ij}-a}{\delta}}$ where $V_{pp\sigma}^0 = t_{\perp}^0 =0.48~\text{eV}$,  $V_{pp\pi}^0 = t_0 =  -2.7~\text{eV}$, $a= 0.142~\text{nm}$, $d_{0} = 0.335~\text{nm}$ and $\delta = 0.184\sqrt{3}a$.  To accurately  describe  the electronic properties of TBG,  for each site $i$, the neighbours $j$ are chosen  inside a disc of radius $d_{ij}\leq 4a$.  To include the effect of the in-plane magnetic field ${\bm B} = B \left (\cos\theta_B {\bm e}_x+ \sin\theta_B {\bm e}_y \right )$ in the tight-binding Hamiltonian, we use the Peierls substitution where the hopping parameters  are modified to  $t_{ij} = t_{ij}e^{i\phi_{ij}}$   where $\phi_{ij} =  ie{\bm A} \cdot \left ({\bm R}_i - {\bm R}_j \right)/\hbar$. In this expression the vector potential ${\bm A} = zB \left (\sin\theta_B{\bm e}_x -\cos\theta_B {\bm e}_y\right)$ is evaluated at $\left ({\bm R}_i + {\bm R}_j \right)/2$. Furthermore, we use a symmetric arrangement where top and bottom layer are located at $z = \pm d/2$ being $ d = 0.335$ nm the interlayer distance.

\begin{center}
\begin{figure}[h!]
\scalebox{0.98}{\includegraphics{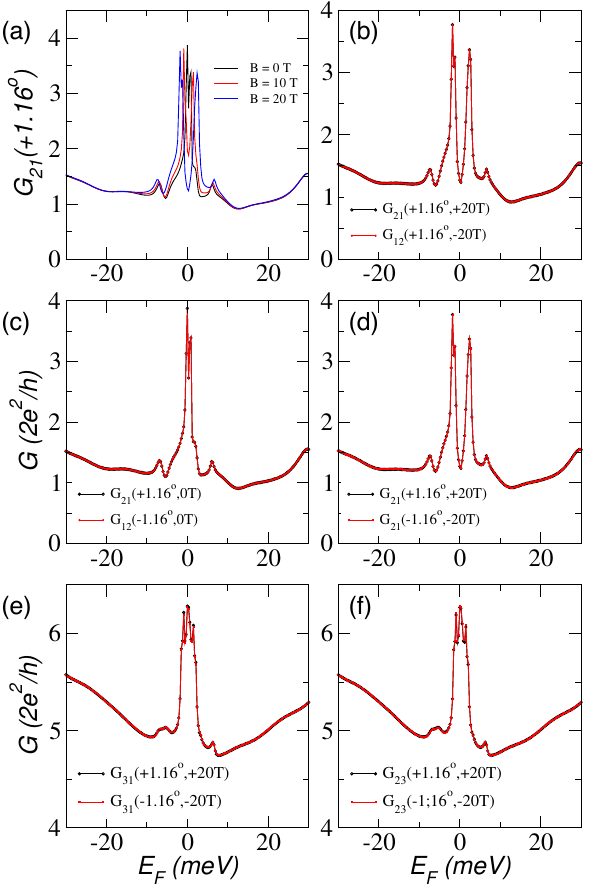}}
\caption{ (Color online) Illustration of field dependence and symmetry properties of conductance matrix entries.  (a) Conductance $G_{21}(\pm 1.16^{\circ},B)$ in units of $2e^2/h$ as a function of the Fermi energy $E_F$, showing the effect of the in-plane magnetic field ${\bm B}=B{\bm e}_x$. (b)  The reciprocity relations, $G_{21}(+{\bm B}) = G_{12}(-{\bm B})$. (c) $G_{21}(+\theta,{\bm B}={\bm 0}) = G_{21}(-\theta,{\bm B}={\bm 0})$. (d)-(f) The chiral reciprocity relations, $G_{pq}(\theta, {\bm B}) = G_{pq}(-\theta, -{\bm B})$.
}
\label{fig:recipro} 
\end{figure}
\end{center}

\subsection{Numerical validation of reciprocity relations}
In the main text, we derived the standard and {\it chiral} reciprocity relation
\begin{align}
\label{eq:recip}
G_{pq}(\theta,{\bm B}) = G_{qp}(\theta, -{\bm B})\;,
\end{align}
\begin{align}
\label{Treflection}
G_{pq}(\theta, {\bm B}) = G_{pq}(-\theta, -{\bm B})\;.
\end{align}
 
We will now numerically validate the reciprocity relations Eqs. (\ref{eq:recip}) and (\ref{Treflection}) in order to show the reliability of our numerical approach. To this end, we distinguish the effect of an in-plane magnetic field on the conductance and then change twist angle and the field orientation.  The black line in Fig. \ref{fig:recipro}(a) corresponds to  $G_{21}$ for a TBG junction with $\theta = +1.16^\circ$ without magnetic field, a peak originated by Van Hove singularities around the CNP is clearly appreciated. \cite{PhysRevB.107.045418,PhysRevLett.132.076302}  When the magnetic field is switched on along the $+x$-direction, the effect is mainly observed for low energies as a  splitting of the conductance peak. This is produced by the separation of the Dirac points \cite{kwan2020twisted, pershoguba2010energy,de2012manipulation} as function of the strength and orientation of the field, that when projected onto the transport direction, appear at different energies (momentum)  of the incoming electrons. Inverting the direction of the field, i.e.,  the field pointing to the negative $x$-direction, we observe that the traditional reciprocity relations, Eq. (\ref{eq:recip}), are fulfilled. In Fig. \ref{fig:recipro}(b), we only present the case $G_{21}(+1.16^\circ,+20~T) = G_{12}(+1.16^\circ,-20~T)$ to keep the discussion more transparent.

Without a magnetic field, it is not possible to detect the handedness of the TBG junction for layer-symmetric leads. To check this numerically, we again select the specific case $G_{21}(+1.16^\circ,0~T) = G_{12}(-1.16^\circ,0~T)$ in Fig. \ref{fig:recipro}(c). However, this relation is no longer valid if an in-plane magnetic field is present as already anticipated. In Figs.  \ref{fig:recipro}(d)-(f), we recognize that the chiral reciprocity relations, Eq. (\ref{Treflection}), are fulfilled, i.e., $G_{21}(+1.16^\circ,+20~T) = G_{21}(-1.16^\circ,-20~T)$,   $G_{31}(+1.16^\circ,+20~T) = G_{31}(-1.16^\circ,-20~T)$ and  $G_{23}(+1.16^\circ,+20~T) = G_{23}(-1.16^\circ,-20~T)$. Although, in Fig. \ref{fig:recipro} we only focus on specific twist angles ($\theta = \pm1.16^\circ$) and fields (${\bm B} = \pm 20{\bm e}_x~T$) the reciprocity relations, Eqs. (\ref{eq:recip}) and (\ref{Treflection}), are valid irrespective of the twist angle and field intensity.

\begin{center}
\begin{figure}[h!]
\scalebox{0.98}{\includegraphics{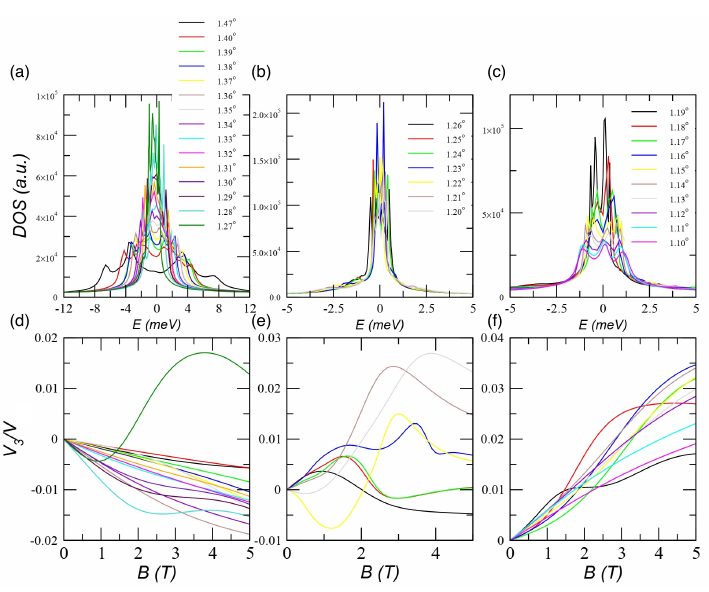}}
\caption{ (Color online) (a)-(c) Density of states for different twist angles without magnetic field. (d)-(f) $V_3/V$ for the angles indicated in the DOS.
}
\label{fig:SIV4DOS} 
\end{figure}
\end{center}

\section{Magic angle regime and density of states}

In the main text, we presented the highest value and the full width half maximum (FWHM) of the density of states (DOS). Based on that, three regimes were identified and we labeled the second regime as magic angle or flat band regime as it displayed the largest peaks and the smallest FWHMs. The twist angles around $1.23^\circ$ are slightly larger than the one found in experiments of $\sim1.1^\circ$ due to the finite system size considered here.  

The DOS was calculated by summing the diagonal elements of the spectral function
\begin{align}
\rho(E) = \frac{1}{2\pi}\text{Tr}[A],
\end{align} 
which is defined as
\begin{align}
A = \mathcal{G}^R(\Gamma_1 + \Gamma_2 + \Gamma_3)\mathcal{G}^A.
\end{align} 
The couplings $\Gamma_{1(2)(3)} = i[\Sigma_{1(2)(3)} - \Sigma_{1(2)(3)}^\dagger]$  and the retarded Green's function  $\mathcal{G} = [E-H_{TBG}-\Sigma_1-\Sigma_2 - \Sigma_3]^{-1}$ were defined in the main text. Note that we neglect the inclusion of an  infinitesimal complex value in  the retarded Green's to avoid any possible effect in the chiral response of the system.\cite{liu2021chirality} This fact has no effect on the transport properties because the self-energies ($\Sigma_i$)  have  their own imaginary part. However, they are  non-zero only for sites in the vicinity of the leads. As a result,  the peaks of the spectral function and consequently the DOS are not perfectly smoothed around the CNP where a large number of states is present.\cite{Datta}  

In Fig. \ref{fig:SIV4DOS}, we present the full  DOS and also the voltage probe $V_3$ for a large range of twist angles between $\theta = 1.47^{\circ}-1.10^{\circ}$. The left  panel shows the results for twist angles $\theta > 1.26^{\circ}$ where negative slopes in the linear regime are found for positive magnetic fields. The central panel presents the DOS and $V_3$ for twist angles around the magic angles ($1.20^{\circ} \le \theta \leq 1.26^{\circ}$); in this regime, no definite slope can be assigned to the voltage probe $V_3$ around $B\approx0$. Finally, the right panel shows the same information for twist angles $\theta < 1.20^{\circ}$, where the slope is reversed and thus positive in the linear regime.  

This result show that there is a change of chirality around the magic angle as found in the extended system.\cite{Stauber20C} The observed voltage probe is thus an indication of an electronic chirality which is not directly linked with the real space chirality which does not change for positive twist angles.

\begin{center}
\begin{figure}[h!]
\scalebox{0.98}{\includegraphics{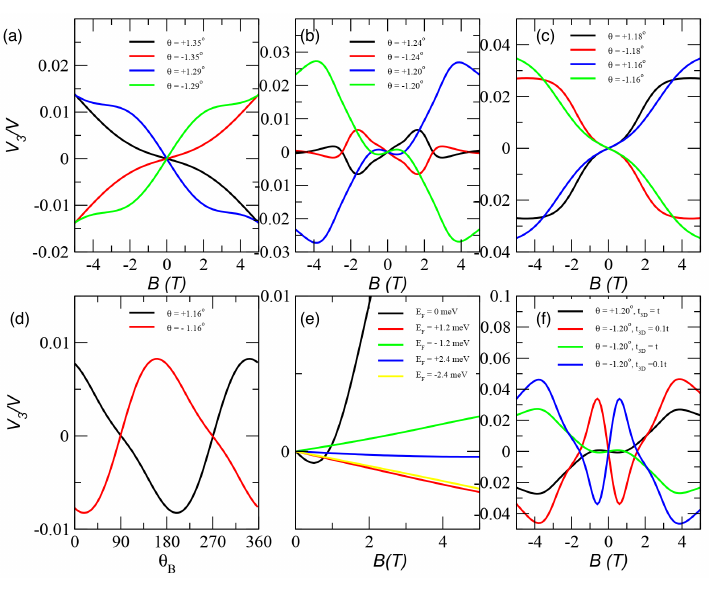}}
\caption{ (Color online) (a)-(c) $V_3$ in units of $V$ for the positive and negative twist angles with $E_F = 0$ and $\bm B$ parallel to the $x$-axis.  (d) Angular dependence of $V_3$, where $\theta_B$ is the angle between the in-plane magnetic field and the $x$-axis, for $\theta = \pm 1.16^{\circ}$ and $|{\bm B}| = 1$ T. (e) Effect of the Fermi energy on $V_3$ for $\theta = 1.20^{\circ}$. (f) $V_3$ changing the coupling between the central region and lead 3 ($t_{3D}=0.1t$).
}
\label{fig:masmenos} 
\end{figure}
\end{center}

\section{Dependence of the voltage probe on magnetic field direction and coupling}

Let us now discuss further results, i.e., the dependence of $V_3$ on negative magnetic fields, on the angle of the magnetic field with respect to the symmetry axis, on a backgate, as well as on the coupling to lead 3. In all cases we choose a symmetric source-drain voltage drop,$V_1=-V_2 =V/2$.

\subsection{Behavior for negative magnetic fields}
 The symmetry arguments of the main text indicate that the relation,  $V_3(+{\bm B},\theta) = -V_3(-{\bm B},\theta)$, is valid to linear order in $\bm B$, in principle. Yet, the results seen in  Figs. \ref{fig:masmenos}(a)-(c) point to an exact relation. This is true because, in those Figs., the field is along the $x$ direction of Fig. 1 of the main text, where an exact symmetry of our geometric setup holds, as explained later when considering non-linear effects. It is also obviously that the behavior is independent of whether the twist angle is above, around, or below the twist angle. 

\subsection{Angle-dependence of the magnetic field on the conductance}

 For the same reason, we have $V_3(+\theta,\bm B) = -V_3(-\theta,\bm B)$ only in the linear field regime, in principle. As before if the magnetic field is along the  $x$ direction, this relation also holds in the non-linear regime. For arbitrary directions,  however, this symmetry is slightly broken as seen in Fig. \ref{fig:masmenos}(d) that shows the angular dependence of $V_3$. In this Figure,  $\theta_B$ denotes the angle between the in-plane magnetic field and the $x$-axis, for $\theta = \pm 1.16^{\circ}$ and $|{\bm B}| = 1$ T. $V_3$ is thus not  symmetric with respect to $\theta_B = \pi$ and the phenomenological correction term  ${\bm I}\cdot {\bm B} = |{\bm I}|| {\bm B}|\cos \theta_B$ that was discussed in a different context for macroscopic devices, see e.g. Refs. \citenum{rikken2001electrical,wagniere2007chirality,NRTokura}, does not match our results. 

\subsection{Gate dependence}
All results so far were obtained for a neutral twisted graphene sample. In Fig \ref{fig:masmenos}(e), we show results also for Fermi energies $E_F = \pm1.2$ and $E_F=\pm2.4$ meV. As can be seen, the linear regime becomes more evident for $\theta = 1.20^{\circ}$ at finite gate voltage. For $ E_F =\pm$1.2meV, we also have $V_3 (E_F)=-V_3(-E_F)$. This might have been expected as also the chiral part of the conductance is an odd function of the chemical potential.\cite{Stauber20C} 

However, this relation is not valid for $E_F = \pm2.4$ meV. We believe this to be a finite size effect as also the DOS of the whole system as function of the twist angle changes, but more detailed simulations are needed in order to understand the gate dependence of the voltage probe. 
\subsection{Chiral engineering} 
Let us investigate the possibility of enhancing the chiral probe in the linear field regime as suggested by Eq. (9) of the main text. In Fig. \ref{fig:masmenos}(f), we show $V_3$ for $\theta = 1.20^{\circ}$ with a reduced coupling between the third lead and the central region ($t_{3D} = 0.1t$). As anticipated, the reduction of the coupling enhances the linear effect showing a possible route to engineer the chiral response.  

\section{Reciprocity relations for the two-terminal conductance}
A dependence of the conductivity on the direction of a magnetic field is forbidden by the general Onsager relations. These state that for any linear response function $\mathcal{K}_{AB}$ one has $\mathcal{K}_{A,B}({\bm B})=\mathcal{K}_{\bar B,\bar A}(-{\bm B})$ with $A,B$ arbitrary operators and $\bar A, \bar B$ the time-reversed operators. For $A=B=j_x$, we have $\bar A=\bar B=-j_x$ and $\mathcal{K}_{j_x,j_x}({\bm B})=\mathcal{K}_{-j_x,-j_x}(-{\bm B})=\mathcal{K}_{j_x,j_x}(-{\bm B})$.

This result also holds within the Landauer formalism in a two-terminal setup, because the unitarity of the S matrix imposes left-right reciprocity for any system attached to
$M_T$ modes, grouped into left and right leads, irrespective of any symmetry (or lack of) of the Hamiltonian. To show this, we first discuss the following relation for the S matrix:
\begin{align}
\sum_{m=1}^{M_T}|s_{mn}|^2=1\;\;\;,\forall n\;,
\end{align}
see Datta (3.1.3b), where $m(n)$ run over all channels attached to our system. Using that $|s_{mn}|^2=|s_{m\leftarrow n}|^2 =T_{m\leftarrow n}$ (Datta (3.1.1)) and grouping these channels into
left(L) and right(R) leads, one has
\begin{align}
\label{UnitaryOne}
\overline{T}_{LL} + \overline{T}_{R\leftarrow L} &= N_L\;,\\
\overline{T}_{RR} + \overline{T}_{L\leftarrow R} &= N_R\;.
\end{align}
where $N_{L(R)}$ is the number of channels in the $L(R)$ lead,
with $N_R + N_L = M_T$ , and
\begin{align}
\overline{T}_{R\leftarrow L} =\sum_{m\in R}\sum_{n\in L} T_{m\leftarrow n}\;,
\end{align}
with similar relations for the other pairings, see Datta
(3.1.2). 

Unitarity also implies the "not so obvious relation"
in Datta's own words, see Datta (3.1.3b) again,
\begin{align}
\sum_{m=1}^{M_T}|s_{nm}|^2=1\;\;\;,\forall n\;,
\end{align}
which leads to
\begin{align}
\label{UnitaryTwo}
\overline{T}_{LL} + \overline{T}_{L\leftarrow R} &= N_L\;,\\
\overline{T}_{RR} + \overline{T}_{R\leftarrow L} &= N_R\;.
\end{align}
The combination of Eqs. (\ref{UnitaryOne}) and (\ref{UnitaryTwo}) implies reciprocity:
\begin{align}
\overline{T}_{R\leftarrow L}=\overline{T}_{L\leftarrow R}\;,
\end{align}
irrespective of any other consideration, except unitarity, always given in the non-dissipative, linear regime.

\section{Relations beyond linear order}

Let us finally discuss extensions to the linear approach presented in the main text. Quite generally, one can always write
\begin{equation}
\begin{bmatrix}   I_1 \\ I_2 \\ I_3   \end{bmatrix} =  (\bm M^{s} + \bm M^{a}) \begin{bmatrix}   V_1 \\ V_2 \\ V_3   \end{bmatrix}\;
,\end{equation}
where $ \bm M^{s}$ is a symmetric matrix, {\em even} in chirality and field, which we can denote as $\bm M^{s}(\theta^2, \bm B^2) $. $ \bm M^{a}$ is an asymmetric matrix that, from unitarity, depends on a single entry
\begin{equation}
\bm M^{a} = \begin{bmatrix}   0 &  -\delta  G  &  +\delta  G \\  +\delta  G & 0  &  -\delta  G  \\ -\delta  G &  +\delta  G & 0 \end{bmatrix}
,\end{equation}
with 
\begin{equation}
\delta G = \frac{G_{12} - G_{21}}{2} = -\frac{G_{13} - G_{31}}{2} = \frac{G_{23} - G_{32}}{2}
,\end{equation}
where now, $\delta G $ is {\em odd} in chirality and field, which we can describe as 
\begin{equation}
\delta G (\theta, \bm B)  = \sigma \sigma' \delta G (\sigma \theta, \sigma' \bm B) 
,\end{equation}
with $\sigma(\sigma')  = \pm 1$. In the main text,  we restricted $\delta G  $ to lowest (linear) order in $ \bm B$, but it can be any odd function of field and chirality, generalizing Eq. (9)  of the main text.

The voltage probe in the case of $I_3=0$ can be written as 
\begin{equation}
V_3  = \frac{V_1 + V_2}{2} +  \delta \tilde{V}_3 \frac{V_1 - V_2}{2},
\end{equation}
with   
\begin{equation}
\delta \tilde{V}_3  = \delta \tilde{V}^{e}_3 + \delta \tilde{V}^{o}_3 ,
\end{equation}
where $\delta \tilde{V}^{e}=\delta \tilde{V}^{e}(\theta^2, \bm B^2)$ is an even function of field and chirality, given explicitly by 
\begin{equation}
\delta \tilde{V}^{e}_3  =  \frac{( G_{31} + G_{13})   - ( G_{32} + G_{23})  }{2 ( G_{31} + G_{32}) }\;.
\end{equation}
It vanishes for $ \bm B = 0 $ in our left-right symmetrically chosen attachment for the third lead as $ G_{31}= G_{32} $. 
 
On the other hand, $\delta \tilde{V}^{o}$, with expression 
\begin{equation}
\delta \tilde{V}^{o}_3  =  \frac{( G_{31} - G_{13})   - ( G_{32} - G_{23})  }{2 ( G_{31} + G_{32}) } = \frac{ \delta G }{( G_{31} + G_{32}) } , 
\end{equation}
is an odd function of field and chirality, formally:
\begin{equation}
  \delta \tilde{V}^{o}(\theta, \bm B) = \sigma \sigma'  \tilde{V}^{o} (\sigma \theta, \sigma' \bm B), 
\end{equation}
with $\sigma(\sigma')  = \pm 1$.  Notice that to linear order in $\bm B$, one has $\delta \tilde{V}_3 (\theta, \bm B)= - \delta \tilde{V}_3 (-\theta, \bm B) $, as asserted in the main text. This relation needs not hold beyond linear order, but, nevertheless, this does not compromise the chiral sensitivity  of the voltage probe for, in general, $ \delta \tilde{V}_3 (\theta, \bm B) \neq \delta \tilde{V}_3 (-\theta, \bm B)$. 

In spite of the above generalizations, our numerical calculations show that, when the field is in the $x$-direction, one always has 
\begin{equation}\label{xaxis}
 \delta \tilde{V}_3 (\theta, B) =  - \delta \tilde{V}_3 (-\theta,  B) , \;  \; \bm B = B {\bm e}_x\;,
\end{equation}
as seen e.g. in Fig. 2 a,b,c. This is no accident, because in our geometry,  a $\pi$ rotation around the $y$-axis reverses $B$ and exchanges leads 1 and 2, leading necessarily to Eq. (\ref{xaxis}) as an exact statement - beyond linear order. This symmetry is also responsible for the vanishing of $\delta \tilde{V}_3 $ when the field is oriented along the $y$-direction, see Fig. 2 d. 

As said, Eq. (\ref{xaxis}) no longer holds beyond linear order when the field is oriented in an  arbitrary direction, as illustrated in Fig. 2 d. Nevertheless,  non-linear effects are rather small, at least for $B=1T$, as can be seen in that figure. Moreover, as previously asserted, the chiral sensitivity of the probe remains such that $ \delta \tilde{V}_3 (\theta, \bm B) \neq \delta \tilde{V}_3 (-\theta, \bm B)$.

\section{Landauer-Büttiker description of the layer-discriminating  setup without magnetic field}
Now we have four leads: 1,  2,  3t, and 3b.  Therefore, 
\begin{equation}\label{MatI}
\left(\begin{matrix} I_1\\I_2\\I_{3t} \\I_{3b}\end{matrix}\right)= \left(\begin{matrix} G_{21}+G_{3t1}+G_{3b1}&-G_{21}&-G_{3t1}&-G_{3b1}\\
                                                                                       -G_{21}&G_{21}+G_{3t1}+G_{3b1}&-G_{3b1}&-G_{3t1}\\
                                                                                       -G_{3t1}&-G_{3b1}&G_{3b3t}+G_{3t1}+G_{3b1} & -G_{3b3t}\\
                                                                                       -G_{3b1}&-G_{3t1}&-G_{3b3t}& G_{3b3t}+G_{3t1}+G_{3b1}\end{matrix}\right)
                                                                                       \left(\begin{matrix} V_1\\V_2\\V_{3t}\\V_{3b}\end{matrix}\right)\;,
\end{equation}
where we have made use of the reciprocity in the absence of magnetic field, $G_{i j} = G_{j i} $, and the additional symmetry of our geometric arrangement, $ G_{3t\, 2} = G_{3b1}$ and $ G_{3t1} = G_{3b2} $. 

\subsection{Detecting chirality with transverse currents: layer contrasting Hall effect}

If the terminals $3t $ and $3b$ are kept at the same potential, $V_{3t}=V_{3b}$, then  
\begin{equation}
\frac{I_{3b}-I_{3t}}{V_1 - V_2} = G_{3t1} - G_{3b1}
,\end{equation}
as asserted in the main text, with a  finite value of this magnitude implying chirality. 

For $V_1=V/2$, $V_2=-V/2$, and  $V_{3b}=V_{3t}=0$, in addition to the standard source-drain current between 1 and 2 reservoirs, one gets transverse currents,
\begin{equation}
I_{3b}+I_{3t} = 0
,\end{equation}
and, therefore, no net current flows from the system to reservoirs $3t + 3b$, kept at the same potential $V_3=0$. Yet a  finite, $|I_{3t}| = |I_{3b}| $, and layer-opposite current in the transverse leads emerges due to chirality, as depicted in Fig. 3 (a) and (b) of the main text. 

The experimental detection of these transverse currents could be problematic, for real leads are always resistive. Nevertheless, the same physics can be exposed with an alternative measurement, where the transverse leads are used as independent voltage probes, and chirality manifests as different voltage readings, $V_{3t} \neq V_{3b}$. This voltage difference can be interpreted physically as the chemical potential difference due to the carrier accumulation-depletion of the "frustrated supercurrent" at the edges. The new conditions would be $V_1=V/2$, $V_2=-V/2$, and  $I_{3b}=I_{3t}=0$.  Eq. \ref{MatI} then gives 
\begin{equation} \label{voltages1} 
 V_{3t} - V_{3b} = \frac{G_{3t 1}-G_{3b 1}}{G_{3t 1}+G_{3b 1} + 2 G_{3b3t}} V 
.\end{equation}
Notice that the denominator of Eq. \ref{voltages1} is even in chirality and, therefore, a finite value of $ (G_{3t\,1} -G_{3b\,1})$,  opposite  for opposite chiralities $(\pm \theta)$, reveals the chirality, as in the case of current detection.

\subsection{Complementary setup for chirality detection}

Guided by results for the conductivity in an infinite system\cite{Stauber18,Stauber18b}, 
%
%
in the main text, we discussed a complementary setup based on the same symmetry properties. It consists of forcing the opposite transverse currents by means of a voltage drop in the corresponding reservoirs, $V_{3t}=+V/2, \; V_{3b}=-V/2$, and then observing as chiral probe the emergence of a net current between reservoirs 1 and 2, without any voltage drop between the terminals, $V_1=V_2=0$,  a "supercurrent" in some sense, see Fig. 3 (c) - (d) of main text.
Under the specified conditions, $V_{3t}=+V/2, \; V_{3b}=-V/2, \, V_1=0, \, V_2=0$, Eq. \ref{MatI} leads to 
\begin{equation}\label{sign}
I_1 = -I_2 = (G_{3b\,1} -G_{3t\,1})V/2,
\end{equation}
that is, the appearance of a net longitudinal current between reservoirs 1 and 2. To avoid potential misunderstandings, notice that the minus sign of $I_2$ in Eq. \ref{sign} comes from usual convention in the  B\"uttiker formalism that a positive  $I_i$ means current {\em injected} from reservoir $i$ to the system, and viceversa. Therefore, Eq. \ref{sign} describes the same amount of current entering to the flake from reservoir 1 and exiting to reservoir 2, again, without any voltage drop between them, $V_1=V_2=0$.

As in the original arrangement of the manuscript, the effect requires chirality manifested as a finite value of $ (G_{3t\,1} -G_{3b\,1})$,  opposite  for opposite chiralities $(\pm \theta)$. The value of $ (G_{3t\,1} -G_{3b\,1})$ as function of Fermi level in twisted bilayer graphene was plotted in the main manuscript.

As before, experimental detection of this "supercurrent" with  real resistive leads can be problematic. We can then opt for using leads 1 and 2 as voltage probes, as in the previous case.  The new conditions would be $V_{3t}=+V/2$, $V_{3b}=-V/2$, and  $I_{1}=I_{2}=0$.  Eq. \ref{MatI} then gives 
\begin{equation} \label{voltages2} 
 V_{1} - V_{2} =  \frac{G_{3t 1}-G_{3b 1}}{G_{3t 1}+G_{3b 1}} \,V 
.\end{equation}
As in Eq. \ref{voltages1},  the denominator of Eq. \ref{voltages2} is  even in chirality and, once again, it is the  finite value of $ (G_{3t\,1} -G_{3b\,1})$,  opposite  for opposite chiralities $(\pm \theta)$, what  reveals the chirality.

\section{Change in chirality in the infinite system with time-reversal symmetry}

\begin{center}
\begin{figure}[!h]
\scalebox{0.5}{\includegraphics{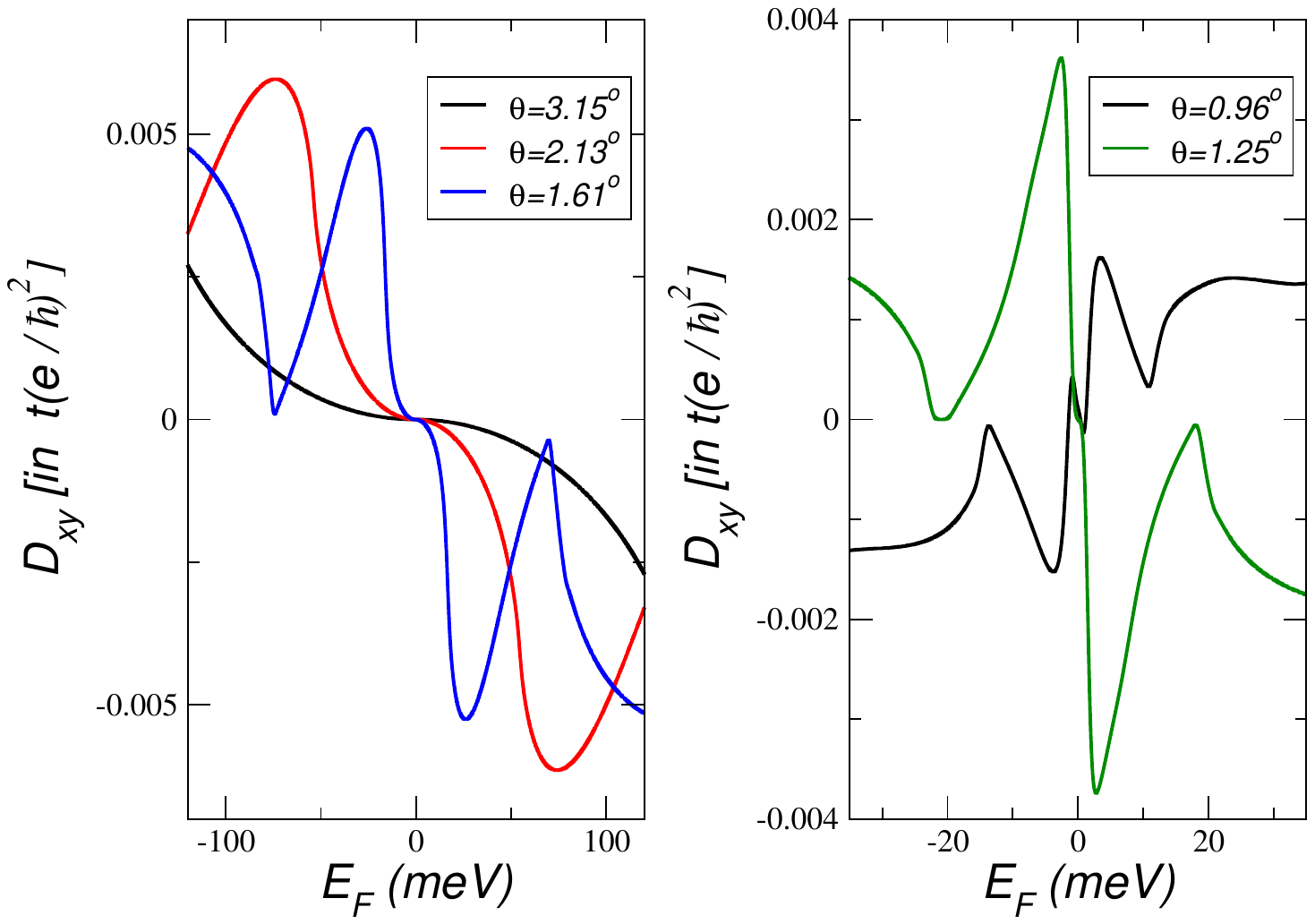}}
\caption{ (Color online) Left panel: The chiral Drude weight for large twist angles. Right panel: The chiral Drude weight below and above the magic angle $\theta_m=1.08^\circ$. 
}
\label{fig:chiralDrude} 
\end{figure}
\end{center}

In Ref. \citenum{Stauber20C}, it was shown that electronic chirality is a Fermi-surface property and given by the chiral Drude weight
\begin{align}
\label{HallDrude}
 D_{xy}=\frac{1}{2A}\sum_{\bm k,n}\bm e_z\cdot(\bm j_{\bm k,n}^1\times\bm j_{\bm k,n}^2)\delta(\epsilon_{\bm k,n}-E_F)\;.
\end{align}
It is defined by the sheet currents of layer $\ell$ denoted by $\bm j_{\bm k,n}^\ell$ and sum over all states labeled by the band index $n$ and Bloch states $\bm k$ with energies around the Fermi energy $E_F$. Furthermore, $A$ denotes the area of the sample and the eigenenergies $\epsilon_{\bm k,n}$ are given in units of the carbon-carbon hopping amplitude $t\sim2.7$eV.

The continuum model of twisted bilayer graphene shows an approximate particle-hole symmetry. Then the chiral Drude weight is in an odd function of the Fermi energy with $D_{xy}(E_F)=-D_{xy}(-E_F)$. The change in chirality at the neutrality point can be seen on the left and right panel of Fig. \ref{fig:chiralDrude}. However, for large angles the chirality is always positive for $E_F<0$. This changes when the twist angle is below the magic angle $\theta_m=1.08^\circ$ and the chirality for e.g. $\theta=0.96^\circ$ is negative for $E_F<0$. 

Interestingly, the shape of the curves is similar to the one of the right panel for positive twist angle of Fig. 4 of the main text. In contrary to the case of the infinite system, these curves do not change around zero due to a residual chirality which is due to finite size effects and not related to the band structure. 

\bibliography{magnetochiralBib}

\end{document}